\documentclass[dvips,12pt]{article}
\usepackage{a4,graphics,psfig,t1enc}
\usepackage{epsf,amsmath,amssymb,latexsym}
\DeclareMathOperator{\rpart}{Re}
\DeclareMathOperator{\impart}{Im}

\begin{document}

\setcounter{page}{1}

{\small{\date{\today}}}
\title{Orbit bifurcations and the scarring of wavefunctions\footnote{Short
 title: Bifurcations and scarring}}
\author{J.~P.~Keating$^{1}$ and S.~D.~Prado$^{2}$ \\
{\small $1$ School of Mathematics, University of Bristol, Bristol} \\ 
{\small BS8 1TW, UK, and BRIMS, Hewlett-Packard Laboratories,} \\
{\small Filton Road, Stoke Gifford, Bristol BS34 8QZ, UK.}\\
{\small $2$ Instituto de F{\'\i}sica, Universidade Federal do Rio
 Grande do Sul} \\
{\small P.O. Box 15051, 91501-970 Porto Alegre, RS, Brazil.}
}
\maketitle
\begin{abstract}
We extend the semiclassical theory of scarring of quantum 
eigenfunctions $\psi_{n}({\bf{q}})$ by classical periodic orbits to 
include situations where these orbits undergo generic bifurcations.
It is shown that $\left |\psi_{n}({\bf{q}}) \right |^{2}$, averaged locally 
with respect to position ${\bf q}$ and the energy spectrum $\{E_{n}\}$, has
structure around bifurcating periodic orbits with an amplitude and 
length-scale whose $\hbar$-dependence is determined by the bifurcation
in question. Specifically, the amplitude scales as $\hbar^\alpha$ and the 
length-scale as $\hbar^\omega$, and values of the {\em scar exponents}, 
$\alpha$ 
and $\omega$, are computed for a variety of generic bifurcations.  In each 
case, the scars are semiclassically wider than those associated with isolated 
and unstable periodic orbits; moreover, their amplitude is 
at least as large, and in 
most cases larger.  In this sense, bifurcations may be said to give rise to 
{\em superscars}.  The competition between the contributions from different
bifurcations to determine the moments of the averaged 
eigenfunction amplitude is
analysed.  We argue that there is a resulting universal $\hbar$-scaling 
in the semiclassical asymptotics of these moments for irregular states in 
systems with a mixed phase-space dynamics.  
Finally, a number of these predictions are illustrated by numerical 
computations for a family of perturbed cat maps. 
\end{abstract}

\newpage

\section{Introduction}

\hspace{\parindent}

One of the main goals in quantum chaology has been to determine the 
link between classical periodic orbits and quantum spectral 
fluctuations in the semiclassical limit. In fully chaotic systems, 
where the periodic orbits are isolated and unstable, this connection
is embodied in Gutzwiller's trace formula (Gutzwiller 1971), and in
integrable systems by a corresponding expression involving the phase-space
tori (Berry \& Tabor 1976). These formulae fail, by diverging, when
periodic orbits bifurcate; that is, when combinations of stable and 
unstable orbits collide and transmute, or annihilate, as a system 
parameter varies --  phenomena that characterize dynamics in the mixed
regime. They must then be replaced by transitional or uniform 
approximations which interpolate through the bifurcation (Ozorio de
Almeida \& Hannay 1987; Tomsovic {\em et al.} 1995; Ullmo {\em et al.} 1996;
 Sieber 1996; 
Schomerus \& Sieber 1997; Schomerus 1998; Sieber \& Schomerus 1998).

That individual orbit bifurcations can have an important, and sometimes
dominant influence on spectral statistics was pointed out by Berry 
{\em et al.} (1998), and demonstrated for a particular example, the 
perturbed cat maps. More generally, Berry {\em et al.} (2000) developed
a semiclassical theory for the competition
between the various generic bifurcations found in Hamiltonian systems
to determine the moments of the quantum energy level counting function. This
suggests that these moments diverge in a universal way, characterized
by certain {\em twinkling exponents}, as $\hbar \rightarrow 0$.

A second major goal of quantum chaology has been to understand the 
influence of classical periodic orbits on quantum wavefunctions in the 
semiclassical limit. It was first noticed by McDonald (McDonald 1983) 
that individual eigenfunctions can have enhanced intensity along short
periodic orbits in classically chaotic systems. This phenomenon was 
later studied systematically by Heller (Heller 1984), who called
such structures {\em scars}. He developed a theory of scarring, based
on wavepacket dynamics, which has subsequently been extended to 
describe a variety of statistical properties of quantum chaotic
eigenfunctions (Kaplan 1999).

An alternative theory of scarring, based on an approach closely related
to the trace formula, was initiated by Bogomolny (Bogomolny 1988). In
this, the semiclassical approximation to the energy-dependent Green
function is used to show that for quantum eigenfunctions
 $\psi_n({\bf q})$  corresponding to energy
levels $E_{n}$ in a fixed energy range, 
$\left < \left | \psi_n({\bf q})\right |^{2} \right >$, where 
$\left < \cdots \right >$ denotes an average over the states in question
and locally over position ${\bf q}$, has complex-Gaussian fringes with, in 
two-degree-of-freedom systems, amplitude and length-scale of
the order of $\hbar^{1/2}$ around unstable periodic orbits. A 
corresponding theory for Wigner functions was developed by Berry (1989).

We emphasize two limitations of the theories mentioned above. First, 
they only describe scarring in eigenfunctions that have been averaged 
over an energy
interval which, semiclassically, contains a large number of states. 
Resummation techniques have been applied to provide
some information about individual eigenfunctions (Agam \& Fishman 1994;
Fishman {\em et al.} 1996), but a detailed understanding of the 
phenomenon in this case remains to be developed. Second, they 
concentrate specifically  on the influence of periodic
orbits which are isolated and unstable. (Semiclassical theories 
describing the connection between quantum wavefunctions and phase-space
tori in classically integrable systems have also been developed; see, 
for example, Berry 1983 for a detailed review.)

Our purpose here is to address the second of these limitations. 
Specifically, our aim is to show how Bogomolny's theory can be extended
to include the description of semiclassical structures in quantum 
eigenfunctions associated with generic classical periodic orbit 
bifurcations in systems with two degrees of freedom. We focus in 
particular on the $\hbar$-dependence of the amplitude and length-scale
of the fringes corresponding to 
those identified by Bogomolny. Our main result is that the amplitude
is of the order of $\hbar^{\alpha}$, and the length-scale is of the order
of $\hbar^{\omega}$, where $\alpha$ and $\omega$ are bifurcation-dependent
{\em scar exponents} whose values we calculate in a number of 
different cases. Crucially, $\omega<1/2$ for all the bifurcations
studied, and for most $\alpha<1/2$ as well. In this sense, 
bifurcations may be said to give rise to {\em superscars}. In order 
to quantify this, we determine the way in which bifurcating orbits 
contribute, via a competition, to the semiclassical asymptotics of the 
moments of 
$\left < \left | \psi_{n}(q)\right |^{2} \right >$, in the same way
as was done for spectral fluctuations by Berry {\em et al.} (2000).  It is 
argued that this competition results in universal $\hbar$-scalings of the 
moments for the irregular eigenfunctions of systems with mixed
phase-space dynamics.
Finally, as an example, we apply some of the techniques developed to 
study the influence of one particular bifurcation on the eigenfunctions
of a family of quantum perturbed cat maps.   

\section{Scar Formulae}

Our aim in this section is to derive semiclassical scar formulae for
bifurcating periodic orbits which generalize those obtained in Bogomolny (1988)
 for unstable orbits far from bifurcation.

We begin, following Bogomolny, with the energy-dependent Green function

\begin{equation}
G({\bf q'},{\bf q};E)=\sum_{n}\frac{\psi^{\ast}_{n}({\bf q'})\psi_{n}({\bf q})}{E-E_{n}},
\label{f1}
\end{equation}
where $\psi_{n}({\bf q})$ is the eigenfunction of the quantum Hamiltonian 
corresponding to the energy level $E_{n}$. The identity we seek to 
exploit follows from setting ${\bf q'}={\bf q}$:

\begin{equation}
\sum_{n}\left | \psi_{n}({\bf q}) \right |^{2}\delta_{\varepsilon}
\left (E-E_{n} \right )=-\frac{1}{\pi}\impart G({\bf q},{\bf q};E+i\varepsilon).
\label{f2}
\end{equation}

\noindent Here, $\delta_{\varepsilon}(x)$ is a normalized, 
Lorentzian-smoothed $\delta$-function of width $\varepsilon$. (It is
straightforward to transform (\ref{f2}) to give differently smoothed
$\delta$-functions, for example Gaussians.) The left-hand side of 
(\ref{f2}) thus corresponds to a sum over eigenstates for which 
$E_{n}$ lies within a range of size of the order of $\varepsilon$ centred on 
$E$. Semiclassically, it is approximately the average of $|\psi_{n}({\bf q})|^{2}$ 
with respect to these states multiplied by $\overline{d}(E)$, the
mean level density. For systems with two-degrees-of-freedom
\begin{equation}
\overline{d}(E) \sim \frac{V(E)}{(2\pi\hbar)^{2}}
\label{f3}
\end{equation}
as $\hbar \rightarrow 0$, where 
\begin{equation}
V(E)=\int \delta (E-H({\bf p},{\bf q})) d^{2}{\bf q} d^{2}{\bf p}
\label{f4}
\end{equation}
and $H({\bf p},{\bf q})$ is the classical Hamiltonian.

The connection with classical mechanics is achieved using the 
semiclassical approximation to the Green function. For systems with 
two-degrees-of-freedom, this is 
\begin{equation}
G({\bf q'},{\bf q};E)\approx \frac{1}{i\hbar \sqrt{2\pi i \hbar}}
\sum_{\gamma}\sqrt{\left | D_{\gamma} \right |} \exp{ \left \{\frac{i}{\hbar}
S_{\gamma}({\bf q'},{\bf q};E)-\frac{i\pi}{2}\nu_{\gamma} \right \}},
\label{f5}
\end{equation}
where the sum includes all classical trajectories from ${\bf q}$ to ${\bf q'}$
at energy $E$, $S_{\gamma}$ is the action along the trajectory labelled
$\gamma$,
\begin{equation}
D_{\gamma}=\det{ \left (
\begin{array}{ll}
\frac{\partial^{2}S_{\gamma}}{\partial {\bf q'}\partial {\bf q}} &
\frac{\partial^{2}S_{\gamma}}{\partial {\bf q'}\partial E} \\
\frac{\partial^{2}S_{\gamma}}{\partial E \partial {\bf q}} &
\frac{\partial^{2}S_{\gamma}}{\partial E^{2}}
\end{array} \right )},
\label{f6}
\end{equation}
and $\nu$ is the Maslov index (Gutzwiller 1990). When ${\bf q'}={\bf q}$,
the sum in (\ref{f5}) is clearly over closed orbits.

We note in passing that it follows from (\ref{f2}) that
\begin{equation}
\sum_{n}\delta_{\varepsilon}(E-E_{n})=-\frac{1}{\pi}\impart \int G({\bf q},{\bf q};E+
i\varepsilon) d^{2}{\bf q}.
\label{f7}
\end{equation}
Substituting in the closed orbit sum for $G({\bf q},{\bf q};E+i\varepsilon)$ and
integrating term-by-term using the method of stationary phase leaves 
contributions from the periodic orbits. Assuming these are all isolated,
as is the case for hyperbolic systems, the result is the trace 
formula (Gutzwiller 1971)

\begin{multline}
\sum_{n}\delta_{\varepsilon}(E-E_{n})\approx \overline{d}(E) +  
\frac{1}{\pi \hbar} \sum_{p}\sum_{r=1}^{\infty} \frac{T_{p}}
{\sqrt{\left | \det{ ({\bf M}^{r}_{p}-{\bf I})}\right |}} \times \\
\cos{\left (
\frac{rS_{p}}{\hbar}- r\nu_p \frac{\pi}{2}\right ) }\exp{ \left [-\varepsilon 
\frac{rT_{p}}
{\hbar}\right ] },
\label{f8}
\end{multline}
where $p$ labels primitive periodic orbits with period $T_{p}$ and
monodromy matrix ${\bf M}_{p}$, and $r$ labels repetitions. As noted
in the Introduction, this formula fails at bifurcations, where
$\det{({\bf M}^{r}_{p}-{\bf I})}=0$. Assuming that the periodic orbits lie in
families which form tori in phase space gives the corresponding
expression for integrable systems (Berry \& Tabor 1976).

Bogomolny's scar formula follows not from integrating over all 
positions ${\bf q}$, as in (\ref{f7}), but from performing a {\em local 
average} of (\ref{f2}) with respect to ${\bf q}$ (we postpone specifying the
size of the averaging range until after the result has been stated), which we
take to be smooth (convolution with a normalized Gaussian, for example). On
the left-hand side this gives, approximately, $\overline{d}(E)\left <
 \left | \psi_{n}({\bf q}) \right |^{2} \right >$, where 
$\left < \cdots \right >$ denotes a combination of the spectral average
described above and the local ${\bf q}$-average.  On the
 right-hand side, the ${\bf q}$-average selects from the closed orbits
those that are close to periodic orbits (i.e. for which the change
in momentum after return is appropriately small). These can then be 
described by linearizing about the periodic orbits. Essentially, this
corresponds to expanding the action up to terms which are quadratic
in the distance from the periodic orbit. The result is that
\begin{multline}
\sum_{n} \left < \left | \psi_{n}({\bf q})\right |^{2}\right >_{\bf q} 
\delta_{\epsilon}(E-E_{n})\approx 
\frac{1}{(2\pi \hbar)^{2}}\Omega({\bf q};E) - \\
\frac{1}{\pi\hbar^{3/2}}\impart 
\frac{1}{i\sqrt{2\pi i}}\sum_{p}\sum_{r=1}^{\infty} 
\frac{1}{\left |\dot{z}\right |
\sqrt{\left [ {\bf M}^{r}_{p} (z) \right ]_{12}}} \times \\
 \exp{
\left [\frac{i}{\hbar} \left ( r S_{p}+\frac{1}{2}
\frac{\det{\left ({\bf M}^{r}_{p}-I \right )}}
{\left [{\bf M}^{r}_{p}(z)\right ]_{12}}y^2-r\nu_{p}\frac{\pi}{2} \right ) 
\right ]} 
\exp{\left [-\varepsilon \frac{rT_{p}}{\hbar}\right ]},
\label{f9}
\end{multline}
where $z$ is a coordinate along a given periodic orbit and $y$ is a
coordinate transverse to it, 
$\left [ {\bf M}^{r}_{p} (z) \right ]_{ij}$ denotes
the elements of the monodromy matrix (which are functions of $z$), 
$\dot{z}$ is the velocity along the periodic orbit, and
\begin{equation}
\Omega({\bf q};E)=\int \delta(E-H({\bf p},{\bf q}))d^{2}{\bf p}.
\label{f10}
\end{equation}
This in turn implies that
\begin{multline}
\left < \left | \psi_{n}({\bf q}) \right |^{2} \right > \approx
\frac{\Omega({\bf q};E)}{V(E)} - \frac{4\pi\sqrt{\hbar}}{V(E)} \impart 
\frac{1}{i\sqrt{2\pi i}}\sum_{p} \sum_{r=1}^{\infty}
\frac{1}{\left | \dot{z}\right |
\sqrt{\left [ {\bf M}^{r}_{p}(z) \right ]_{12}}} \times \\
\exp{
 \left [
 \frac{i}{\hbar} \left (
 r S_{p} +\frac{1}{2}\frac{\det{\left ( 
{\bf M}^{r}_{p}-{\bf I} \right )}}
{\left [ {\bf M}^{r}_{p}(z) \right ]_{12}}
y^2-r\nu_{p}\frac{\pi}{2}\right )
 \right ] }   
 \exp{\left [ -\varepsilon \frac{ r T_{p}}{\hbar} \right ]  }.
\label{f11}
\end{multline}

Equation (\ref{f11}) is Bogomolny's scar formula. In classically ergodic
systems, the first term represents
the quantum-ergodic limit of the eigenfunction 
probability density (Shnirelman 1974, 
Colin de Verdi\`ere 1985, Zelditch 1987).  The second  describes complex
Gaussian fringes (the $y$-dependent part), with length-scale and 
amplitude both of the order of $\hbar^{1/2}$, associated with each 
periodic orbit. This structure will be resolved if the local 
${\bf q}$-average is over regions whose dimensions 
are small compared to the length-scale
of the fringes; that is, over regions whose dimensions scale 
as $\hbar^{\delta}$,
where $\delta > 1/2$. In order for the near-to-periodic-orbit 
approximation to be valid, we must also have $\delta<1$; that is, the 
dimensions of the 
averaging range must be large compared to a de Broglie wavelength.

The trace formula (\ref{f8}) can be recovered from (\ref{f9}) by 
integrating over $z$ and $y$. The $z$-integral gives the period in the
amplitude of the periodic orbit contributions, and the $y$-integral
gives the determinant. Note that the power of $\hbar$ in the trace
formula amplitude is the amplitude exponent in (\ref{f9}), $-3/2$
(which in turn is equal to the amplitude exponent in (\ref{f11}) minus
two -- the exponent in $\overline{d}$), plus the length-scale exponent
of the fringes, $1/2$.

The approximations (\ref{f9}) and (\ref{f11}) break down in two ways. First,
at self-focal points along an orbit 
$\left [ {\bf M}^{r}_{p} \right (z)]_{12}=0$ and the amplitude diverges. 
This can be remedied straightforwardly using Maslov's method, and we 
will not concern 
ourselves further with it here. Second, when an orbit bifurcates
$\det{({\bf M}^{r}_{p}-{\bf I})}=0$, and so the formulae become
$y$-independent. Essentially, this means that the fringes are 
infinitely wide (it is this infinity which, upon integration with 
respect to $y$, transfers itself to the amplitude in the trace 
formula). Our purpose in this paper is to show how to correct (\ref{f9})
and (\ref{f11}) in this case. 

It might be thought that the scar formulae
for bifurcating orbits could be obtained easily by expanding the 
action in (\ref{f5}) to higher order than quadratic. For some of the 
simpler bifurcations (e.g. the codimension-one bifurcations of orbits 
with $r=1$ and $r=2$) this is correct (see the example in Section 4).
However, for more complicated bifurcations it is incorrect, because
for these the linearized map ${\bf M}^{r}_{p}$ is equal to the identity,
which cannot be generated by the action $S({\bf q'},{\bf q};E)$. Thus it is
difficult to build into the semiclassical expression for the 
${\bf qq'}$-representation of the Green function a well-behaved description of
the nonlinear dynamics which the linearized map approximates. The
solution to this problem, originally proposed in Ozorio de Almeida
\& Hannay (1987), is to transform the Green function to a mixed
position-momentum representation, and this is the approach we now take.
     
The Ozorio de Almeida-Hannay method involves, first, Fourier 
transforming $G({\bf q'},{\bf q};E)$ with respect to ${\bf q'}$. 
This gives the 
Green function in the ${\bf q}$ ${\bf p'}$-representation, 
$\tilde{G}({\bf p'},{\bf q};E)$
(${\bf p'}$ is the momentum conjugate to ${\bf q'}$). The semiclassical 
approximation to $\tilde{G}$ takes the same form as (\ref{f5}),
except that $S({\bf q'},{\bf q};E)$ is replaced by the 
${\bf q}$ ${\bf p'}$-generating 
function $\tilde{S}({\bf p'},{\bf q};E)$. $G({\bf q'},{\bf q};E)$ 
may then be rewritten,
semiclassically, as the Fourier transform of this expression with 
respect to ${\bf p'}$. (For an alternative approach leading to the same
final answer, see Sieber 1996). The result, for the semiclassical 
contribution to $G({\bf q},{\bf q};E)$ from closed orbits in the neighbourhood
of a bifurcating periodic orbit, takes the following form.

Consider the case of a codimension-$K$ bifurcation of a periodic 
orbit with repetition number $r$. 
As before, let $z$ be a coordinate along the orbit at
bifurcation, let $y$ be a coordinate transverse to it, and let $p_{y}$
be the momentum conjugate to $y$, so that $y$ and $p_{y}$ are local
surface of section coordinates. Let $\Phi_{r,K}(y,  p_{y},{\bf x})$
be the normal form which corresponds to the local (reduced) generating
function in the neighbourhood of the bifurcation (Arnold 1978;
Ozorio de Almeida 1988), where ${\bf x}=(x_{1},x_{2},\cdots,
x_{K})$
are parameters controlling the unfolding of the bifurcation.
Then, up to irrelevant factors, the contribution to $G({\bf q},{\bf q};E)$ is
\begin{equation}
G_{r,K}(y;{\bf x})=\frac{1}{\hbar^2}\int 
\exp{\left [\frac{i}{\hbar}
\Phi_{r,K}(y,p_{y},{\bf x})\right ]} dp_{y}
\label{f12}
\end{equation}
(as already stated, we are here interested in determining the
$\hbar$-dependence of the amplitude and length-scale of the associated
fringes, and so have neglected terms in (\ref{f12}), such as an 
$\hbar$-independent factor in the integrand, which do not influence
these). 

Before proceeding further, we make three remarks about (\ref{f12}). 
First, the power of $\hbar$ outside the exponential arises from adding
$1/2$, which comes from the Fourier transform, to the exponent in 
(\ref{f5}), $3/2$. Second, the representation used in Berry {\em et
al.} (2000)
for the fluctuating part of the spectral counting function may be
derived from (\ref{f12}) by taking the trace of $G_{r,K}$, which
involves
 integrating the right-hand side of (\ref{f12}) with respect to 
$y$ (the $z$-integral is trivial, as before), and then integrating
with
respect to $E$ resulting in a further multiplication by $\hbar$.  
Likewise, the formulae of Ozorio de Almeida and Hannay (1987) 
correspond to taking the trace of $G_{r,K}$, keeping the terms we
have neglected. Third, the $\hbar$-dependence of the fringes in 
Bogomolny's scar formula for unstable periodic orbits far from
bifurcation can be recovered using the appropriate normal form:
\begin{equation}
\Phi_{r,0}=p^{2}_{y}+y^{2},
\label{f13}
\end{equation}
which corresponds to a particular, $\hbar$-independent choice of units
for $p_{y}$ and $y$. Evaluating the integral then gives
\begin{equation}
G_{r,0}({\bf q})\propto \frac{1}{\hbar^{3/2}} \exp{\left (
i\frac{y^2}{\hbar} \right )},
\label{f14}
\end{equation}
as in (\ref{f9}).

Equation (\ref{f12}) is the starting point for the analysis of 
bifurcating orbits. Our strategy is essentially the same as that used
in Berry {\em et al.} (2000) to study the related fluctuations in the 
spectral counting function (see also Berry 2000 for a review of 
applications to other areas in wave physics): first, rescale $y$ and
$p_y$ to remove the $1/\hbar$ factor from the 
dominant term (germ) of $\Phi_{r,K}$ in the exponent, and then apply a 
compensating rescaling of the parameters $x_{1},x_{2},\cdots, x_{K}$
to remove the $\hbar$-dependence from the other terms which do not
vanish as $\hbar \rightarrow 0$. This  will lead to 
\begin{equation}
G_{r,K}(y;{\bf x};\hbar)=\frac{1}{\hbar^{2-\alpha_{r,K}}}G_{r,K}
\left (\frac{y}{\hbar^{\omega_{r,K}}},\left
\{ \frac{x_{n}}{\hbar^{\sigma_{n},r,K}}\right \},1 \right ).
\label{f15}
\end{equation}
The exponent $\alpha$ describes the semiclassical amplitude of the fringes in
$\left < \left | \psi_{n}({\bf q}) \right |^{2} \right >$ associated with
the 
bifurcation, and the exponent $\omega$ describes the
$\hbar$-dependence
 of their length-scale, or width. We call these the {\em scar
exponents}. Note that the corresponding amplitude exponent in
 $\sum_{n}\left < \left | \psi_{n}({\bf q}) \right |^{2}\right
>_{q}\delta_\varepsilon (E-E_{n})$ is $2-\alpha$. The exponents $\sigma$
describe the range of influence of the bifurcation in the different
unfolding directions $x_{n}$. Their sum
\begin{equation}
\gamma_{r,K}=\sum_{n=1}^{K}\sigma_{n,r,K}
\label{f16}
\end{equation}
describes the $\hbar$-scaling of the $K$-dimensional
${\bf x}$-space
hypervolume affected by the bifurcation.

We now calculate these exponents in a variety of examples. Consider
first the $r=1$ bifurcations which correspond to cuspoid (i.e. corank
1) catastrophes. For these, the normal forms are (Berry {\em et al.}
2000)
\begin{equation}
\Phi_{1,K}(y,p_{y};{\bf x})=p^{2}_{y}+y^{K+2}+\sum_{n=1}^{K}
x_{n}y^{n}.
\label{f17}
\end{equation}
Substituting this into (\ref{f12}) and evaluating the integral then
gives
\begin{equation}
G_{1,K}(y;{\bf x}) \propto \frac{1}{\hbar^{3/2}}\exp{\left [\frac{i}{\hbar}
\left (
y^{K+2}+\sum^{K}_{n=1}x_{n}y^{n} \right ) \right]} 
\label{f18}
\end{equation}
Making the rescalings $\tilde{y}=y/\hbar^{1/(K+2)}$, 
$\tilde{x}_{n}=x_{n}/\hbar^{1-n/(K+2)}$ removes the $\hbar$-dependence
of the exponent, and so for the cuspoids we have
\begin{equation}
\alpha_{1,K}=\frac{1}{2}, {\mbox{\hspace{1cm}}}  \omega_{1,K}=\frac{1}{K+2}, 
{\mbox{\hspace{1cm}}} 
\sigma_{n,1,K}=1-\frac{n}{K+1}
\label{f19}
\end{equation}
and
\begin{equation}
\gamma_{1,K}=\frac{K(K+3)}{2(K+2)}.
\label{f20}
\end{equation}
Analogous expressions can be written down for the $r=1$ bifurcations
corresponding to the more complicated case of catastrophes of corank 
$2$ (see, for example, Berry 2000) in the same way.

When $r>1$, the generic bifurcations with $K=1$ have been classified
by Meyer (Meyer 1970, 1986; Arnold 1978; Ozorio de Almeida 1988), and
those with $K=2$ by Schomerus (1998). The relevant parts of the 
corresponding normal forms, taken from Berry (2000) (to which readers
are referred for further details), are summarized in Table 1 (in the
expressions given, we are retaining only those terms which affect the
exponents we seek to calculate).

\begin{table}[htb] 
\begin{center}
\begin{tabular}{|c|c|} \hline\hline
$r$ & $\Phi_{r,2}$ \\ \hline
$2$ & $p^{2}_{y}+y^6+x_{1}y^2+x_{2}y^{4}$ \\ 
$3$ & $(p^{2}_{y}+y^{2})^2+x_{1}(p^{2}_{y}+y^{2})+x_{2}
\rpart{\left[ (p_{y}+ i y)^3\right ]}$ \\ 
$4$ & $p^{2}_{y}y^{2}+x_{1}(p^{2}_{y}+y^{2}) + 
x_{2}(p^{2}_{y}-y^{2})^2$\\ 
$5$ & $\rpart{\left [(p_{y}+i
y)^{5}\right ] }+x_{1}(p^{2}_{y}+y^2)+x_{2}(p^{2}_{y}+y^{2})^{2}$ \\ 
$\geq 6$ &
$(p^{2}_{y}+y^{2})^{3}+x_{1}(p^{2}_{y}+y^{2})+x_{2}(p^{2}_{y}+
y^{2})^{2}$ \\ \hline\hline
\end{tabular}
\end{center}
\caption{The relevant parts of the normal forms for $K=2$ bifurcations
of period-$r$ orbits (taken from Berry 2000). The corresponding 
expressions for $K=1$ bifurcations, $\Phi_{r,1}$, follow from settling
$x_{2}=1$.}
\label{table1}
\end{table}

The corresponding scar exponents, and the hypervolume exponents
$\gamma$
are given in Table 2 $(K=1)$ and Table 3 ($K=2$).

\begin{table}[htb]
\begin{center}
\begin{tabular}{|c|c|c|c|} \hline\hline 
$r$      & $\alpha_{r,1}$ & $\omega_{r,1}$ & $\gamma_{r,1}$ \\ \hline 
$2$      & $1/2$     &  $ 1/4$        &     $ 1/2$     \\ 
$3$      & $1/3$     &  $ 1/3$        &     $ 1/3$     \\ 
$\geq 4$ & $1/4$     &  $ 1/4$        &     $ 1/2$     \\ \hline\hline
\end{tabular}
\end{center}
\caption{Scar exponents for generic, codimension-1 bifurcations.}
\label{table2}
\end{table}

\begin{table}[htb]
\begin{center}
\begin{tabular}{|c|c|c|c|} \hline\hline
$r$      & $\alpha_{r,2}$ & $\omega_{r,2}$ & $\gamma_{r,2}$ \\ \hline
$2$      & $1/2$     &   $1/6$        &      $1$       \\ 
$3$      & $1/4$     &   $1/4$        &      $3/4$     \\ 
$4$      & $1/4$     &   $1/4$        &      $1/2$     \\ 
$5$      & $1/5$     &   $1/5$        &      $4/5$     \\ 
$\geq 6$ & $1/6$     &   $1/6$        &      $1$       \\ \hline\hline 
\end{tabular}
\end{center}
\caption{Scar exponents for generic, codimension-2 bifurcations.}
\label{table3}
\end{table}

Finally, we consider bifurcations of orbits for which $r\geq 2K+2$.
In this case, the relevant terms in the normal forms are (Berry {\em
et
al.} 2000)
\begin{equation}
\Phi_{r,K}(y,p_{y};{\bf x})=I^{K+1}+\sum^{K}_{n=1}x_{n}I^{n}+
{\cal{O}}(I^{K+2}),
\label{f21}
\end{equation}
where
\begin{equation}
I=y^2+p^{2}_{y}.
\label{f22}
\end{equation}
Expressing $\Phi$ in terms of $y$ and $p_{y}$, we find
\begin{equation}
\alpha_{r,K}=\omega_{r,K}=\frac{1}{2(K+1)}
\label{f21a}
\end{equation}
and 
\begin{equation}
\gamma_{r,K}=\frac{K}{2}
\label{f21b}
\end{equation}
(c.f. the $r\geq 4$ exponents in Table 2, and the $r\geq 6$ exponents in Table
3).

The main point we wish to draw attention to is that, in all the cases
listed above, $\omega <1/2$ and $\alpha\leq 1/2$, and that in most cases
$\alpha<1/2$. Recall that $\omega=\alpha=1/2$ for periodic orbits far from
bifurcation. In this sense, bifurcations may be said to give rise to
{\em superscars}; that is, to scars that are semiclassically wider,
and
 often greater in amplitude than those associated with non-bifurcating
orbits. We shall demonstrate this with an explicit example in Section
4.

The width exponent $\omega$ determines the scale for the ${\bf q}$-average in
$\left < \left | \psi_{n}({\bf q}) \right |^{2} \right >$ which allows the
fringe structure to be resolved.  Specifically, if the dimensions of the 
averaging range
scale semiclassically as $\hbar^\delta$, the fringe structure will be 
resolved if $\delta > \omega$, but not if $\delta < \omega$.  Recall 
that for the approximation to hold in which the main contribution to the
average comes from closed orbits in the neighbourood of periodic orbits, we 
must also have $\delta<1$; that is, the average must extend over many de 
Broglie wavelengths.

We also note that the scar exponents satisfy
\begin{equation}
\beta=1-\alpha-\omega,
\label{f24}
\end{equation}
where $\beta$ is the amplitude exponent of the fluctuations in the 
spectral counting function associated with the bifurcation in
question
(Berry {\em et al.} 2000). This follows from a comparison of
(\ref{f12}) with the corresponding expression for the counting
function, which, as already noted, corresponds to integrating 
(\ref{f12}) with respect to $y$ and multiplying by $\hbar$. It
generalizes the connection discussed above between the power of 
$\hbar$ in the trace formula for $d(E)$ and the amplitude and width
exponents of Bogomolny's fringes for non-bifurcating orbits.

\section{Moment asymptotics}

One way to quantify scarring effects is in terms of the moments of the 
wavefunctions.  Consider the case when all periodic orbits are isolated and unstable.  We define
\begin{equation}
C_{2m}(\hbar)=\frac{1}{\Delta q}\int \left (
\left < \left | \psi_{n}({\bf q'}) 
\right |^{2} \right >-\frac{\Omega({\bf q'};E)}{V(E)} \right)^{2m}d^2{\bf q'}
\label{f26}
\end{equation}
where the ${\bf q'}$-integral is over an $\hbar$-independent volume 
$\Delta q$.  Note that these are the moments not of the
amplitude of the wavefunction itself, but of the amplitude averaged with
respect to position (over a region which shrinks as $\hbar \rightarrow 0$, 
but which contains an increasing number of de Broglie wavelengths) and 
energy (semiclassically many levels).  

Assuming that the wavefunctions are quantum ergodic on the scale of the
local ${\bf q}$-average implies that $C_{2m}(\hbar) \rightarrow 0$ as 
$\hbar \rightarrow 0$ when $m>0$.  It follows from the fact that the 
fringes in (\ref{f11}) have scar 
exponents $\alpha=\omega=1/2$ that their individual contributions 
to the moments scale as
$\hbar^{m+1/2}$ in this limit.  The corresponding contribution from a 
bifurcating orbit is of the order of $\hbar^{2m\alpha + \omega}$, and so is 
semiclassically larger.  
For the bifurcations of periodic orbits with $r=1$ corresponding to the
cuspoid catastrophes we have that
\begin{equation}
{\tilde{\mu}}_{m,1,K}=2m\alpha_{1,K}+\omega_{1,K}=m+\frac{1}{K+2}.
\label{f26a}
\end{equation}
The 
values of these exponents for the generic bifurcations with $K=1$  and $K=2$ 
when $r>1$ are given in Table 4 and Table 5, for $m \le 3$.

\begin{table}[htb] 
\begin{center}
\begin{tabular}{|c|c|c|c|c|c|c|} \hline\hline
$r$ & ${\tilde{\mu}}_{1,r,1}$ & ${\overline{\mu}}_{1,r,1}$
    & ${\tilde{\mu}}_{2,r,1}$ & ${\overline{\mu}}_{2,r,1}$
    & ${\tilde{\mu}}_{3,r,1}$ & ${\overline{\mu}}_{3,r,1}$  \\ \hline
$2$ & $5/4$ & $7/4$ & $9/4$ & $11/4$ & $13/4$ & $15/4$  \\ 
$3$ & $1$   & $4/3$ & $5/3$ &  $6/3$ &  $7/3$ &  $8/3$  \\  
$\geq 4$    & $3/4$ & $5/4$ &  $5/4$ &  $7/4$ &  $7/4$ & $9/4$ \\
    \hline\hline
\end{tabular}
\end{center}
\caption{Values of ${\tilde{\mu}}_{m,r,1}=2m\alpha_{r,1}+\omega_{r,1}$
and ${\overline{\mu}}_{m,r,1}=2m \alpha_{r,1}+\omega_{r,1}+\gamma_{r,1}$
for the codimension-1 scar exponents listed in Table 2.}
\label{table4}
\end{table}

\begin{table}[htb] 
\begin{center}
\begin{tabular}{|c|c|c|c|c|c|c|} \hline\hline
$r$ & ${\tilde{\mu}}_{1,r,2}$ & ${\overline{\mu}}_{1,r,2}$
    & ${\tilde{\mu}}_{2,r,2}$ & ${\overline{\mu}}_{2,r,2}$
    & ${\tilde{\mu}}_{3,r,2}$ & ${\overline{\mu}}_{3,r,2}$  \\ \hline
$2$ & $7/6$ & $13/6$ & $13/6$ & $19/6$ & $19/6$ & $25/6$  \\ 
$3$ & $3/4$ &  $3/2$ & $5/4$ &     $2$ &  $7/4$ &  $5/2$  \\  
$4$ & $3/4$ & $5/4$ &  $5/4$ &  $7/4$ &  $7/4$ & $9/4$ \\
$5$ & $3/5$ & $7/5$ &    $1$ &  $9/5$ &  $7/5$ & $11/5$ \\
$\geq 6$ & $1/2$ & $3/2$ &  $5/6$ &  $11/6$ &  $7/6$ & $13/6$ \\ \hline\hline
\end{tabular}
\end{center}
\caption{Values of ${\tilde{\mu}}_{m,r,2}=2m\alpha_{r,2}+\omega_{r,2}$
and ${\overline{\mu}}_{m,r,2}=2m \alpha_{r,2}+\omega_{r,2}+\gamma_{r,2}$
for the codimension-2 scar exponents listed in Table 3.}
\label{table5}
\end{table}

The moments defined by (\ref{f26}) are implicitly functions of the system
parameters ${\bf x}$.  Averaging them with respect to ${\bf x}$ produces an
opportunity for a competition between the various generic bifurcations.  The
contribution of each bifurcation must be weighted by the associated hypervolume
in ${\bf x}$-space, and so scales as $\hbar^{2m\alpha + \omega + \gamma}$,
provided that the ${\bf x}$-average of the $2m$th power of the 
$\hbar$-independent term in (\ref{f15}) exists (see 
Berry 1977 for a discussion of this subtle point).
For the $r=1$ bifurcations corresponding to the cuspoid catastrophes,
\begin{equation}
{\overline{\mu}}_{m,1,K}=2m \alpha_{1,K}+\omega_{1,K}+\gamma_{1,K}=
m+\frac{K+1}{2}.
\label{f27a}
\end{equation}
When $r>1$, the values of these exponents for the generic 
bifurcations with $K=1$ and $K=2$
are also listed in Table 4 and Table 5.  
The bifurcation that wins the competition, and hence which determines the 
rate at which the ${\bf x}$-averaged moments tend to zero in the 
semiclassical limit, is the one for which $2m \alpha +\omega + \gamma$ is
minimized; that is,
\begin{equation}
\frac{1}{\hbar^{\mu_m}}\left< 
C_{2m}(\hbar) \right >_{\bf x}=o(\hbar^\epsilon)
\label{f27}
\end{equation}
for any $\epsilon < 0$, with
\begin{equation}
\mu_m={\rm min}(2m\alpha+\omega+\gamma)
\label{f28}
\end{equation}
where $\left < C_{2m}(\hbar) \right >_{\bf x}$ denotes the ${\bf x}$-averaged
moments and the minimum is with respect to the generic bifurcations.  This,
of course, assumes that the minimum exists.  We now argue that it does.

Our reasoning is based directly on that of Berry {\em et al.} (2000), where
the analogous problem of the moments of fluctuations in the level counting 
function was considered.  First, we note that if the competition is 
restricted to bifurcations with $r\geq 2K+2$, then
\begin{equation}
{\rm min}(2m\alpha+\omega+\gamma)={\rm min}_K 
\left (\frac{2m+1}{2(K+1)}+\frac{K}{2} \right ),
\label{f29}
\end{equation}
which exists for any $m$ and can be calculated straightforwardly.  
Second, it was
shown by Berry {\em at al.} that each bifurcation with $r\geq 2K+2$ has a
counterpart with $r < 2K+2$ with the property that the counterpart has 
a normal form with the 
same germ, and so the same $\alpha$ and $\omega$ exponents, but a larger
$\gamma$ exponent.  Hence (\ref{f29}) represents the minimum with respect to
all of the generic bifurcations and so
\begin{equation}
\mu_m={\rm min}_K \left (\frac{2m+1}{2(K+1)}+\frac{K}{2} \right ).
\label{f30}
\end{equation}
For example, $\mu_1=5/4$ (coming from $K=1$), $\mu_2=7/4$ (also coming
from $K=1$), and $\mu_3=13/6$ ($K=2$).  In general $\mu_m \approx 
\sqrt{2m+1}-1/2$.

It is natural to compare $\mu_m$ to the corresponding exponent for Gaussian
random functions, which are often taken as models of quantum chaotic 
wavefunctions.  In that case, the moments (\ref{f26}) are semiclassically
of the order of $\hbar^m$.  This follows from the results of Section IIIB
of Eckhardt {\em et al.} 1995, if the operator considered there is the 
characteristic function of the region over which the local ${\bf q}$-average
in (\ref{f26}) extends.  The same rate of vanishing also holds for the 
eigenvectors of random hermitian matrices (see Section IIIA of Eckhardt 
{\em et al.}).  (Readers are referred to B\"acker {\em et al.} 1998 for a 
detailed review of the rate of quantum ergodicity, and its characterization
by moments analogous to those defined by (\ref{f26})).  Crucially, we note that
$\mu_m \le m$ for $m>1$, and so then, if the background to the scars due to 
individual periodic orbits is modelled by a Gaussian random function, 
bifurcations 
dominate the semiclassical asymptotics.  The contributions from individual
non-bifurcating orbits are always subdominant.

To summarize, the exponents $\mu_m$, which are analogous to the {\em twinkling exponents} 
of Berry {\em et al} (2000), determine the asymptotic scaling of the
parameter-averaged moments $C_{2m}$ in the limit as $\hbar \rightarrow
0$ when $m>1$.  Note that they are universal, that is, system independent.  Note also
that they are determined solely by generic bifurcation processes.  As pointed
out in the Introduction, these processes are characteristic of mixed
phases-space dynamics, and so one might expect the
exponents (\ref{f30}) 
to describe the semiclassical deviations of the irregular (in 
the sense of Percival 1973) eigenfunctions in mixed systems 
from their ergodic limit.  
(They do not describe the regular eigenfunctions, for which
the corresponding moments have a different origin, and can be calculated 
using the results of Berry {\em et al.} 1983.)

\section{Perturbed cat maps}

We now illustrate some of the general ideas described in the previous sections 
by focusing on a particular example: a family of perturbed cat maps.

The maps we consider are of the form
\begin{equation}
\left ( \begin{array}{l} 
         q_{n+1} \\
         p_{n+1} 
        \end{array} \right )=
\left ( \begin{array}{ll}
         2 & 1 \\
         3 & 2
        \end{array} \right ) \left ( \begin{array}{l}
                                      q_n \\
                                      p_{n}
                                     \end{array} \right )+
\frac{\kappa}{2 \pi} \cos(2\pi q_{n}) \left (\begin{array}{l}
                                  1 \\
                                  2
                                 \end{array} \right ) \mbox{mod 1},
\label{f40}
\end{equation}
where $q$ and $p$ are coordinates on the unit two-torus, 
and are taken to be a position and its conjugate momentum.  
These maps are Anosov systems for $\kappa \leq \kappa_{\rm 
max}=(\sqrt{3}-1)/\sqrt{5}\approx 0.33$; that is, for 
$\kappa$ in this range they are completely hyperbolic and 
their orbits are conjugate to those of the map with 
$\kappa=0$ (i.e. there are no bifurcations).  Outside this range, 
bifurcations occur, stable islands are created, and the dynamics becomes mixed 
(see, for example, Berry {\em et al.} 1998, where these systems were used 
to demonstrate the influence of periodic orbit bifurcations on 
long-range spectral statistics).

The quantization of maps like (\ref{f40}) was developed by 
Hannay \& Berry (1980), when $\kappa=0$, and Bas\'ilio de 
Matos \& Ozorio de Almeida (1995) for non-zero $\kappa$.  
The quantum kinematics associated with a phase space that 
has the topology of a two-torus restricts Planck's constant 
to taking inverse integer values.  The integer in question, 
$N$, is the dimension of the Hilbert space of admissible 
wavefunctions.  With doubly  periodic boundary conditions 
(see, for example, Keating {\em et al.} 1999), these 
wavefunctions in their position representation have support 
at points $q=Q/N$, where $Q$ takes integer values between 1 
and $N$.  They may thus be represented by $N$-vectors with 
complex components.  The quantum dynamics is then generated 
by an $N \times N$ unitary matrix ${\bf U}$ whose action on 
the wavefunctions reduces to (\ref{f40}) in the classical 
limit; for example
\begin{equation}
U_{Q_1,Q_2}=\frac{1}{\sqrt{iN}}\exp \left [ \frac{2\pi i} 
{N} (Q_1^2-Q_1Q_2+Q_2^2)+\frac{iN}{2\pi}\kappa\sin (2\pi Q_1/N) \right ].
\label{f41}
\end{equation}
This matrix plays the role of the Green function of the 
time-dependent Shr\"odinger equation for flows.  

Denoting the eigenvalues of ${\bf U}$ by 
${\rm e}^{i\theta_n}$, and the corresponding eigenfunctions by 
$\Psi_n(Q)$, we have that
\begin{equation}
\sum_{n}\left |\Psi_{n}(Q)
\right |^{2}\delta_{\varepsilon}(\theta-\theta_{n}) 
=1+\rpart \sum^{\infty}_{k=1}U_{Q,Q}^{k}\exp{(-i\theta k 
-\varepsilon k)}
\label{f42}
\end{equation}
where
\begin{equation}
\delta_{\varepsilon}(x)=\frac{1-e^{-\varepsilon}\cos{x}}{1+e^{-2\varepsilon}-2e^{-\varepsilon}\cos{x}}
\label{f43}
\end{equation}
is a periodized, Lorentzian-smoothed $\delta$-function of 
width $\varepsilon$ (Keating 1991).  Equation (\ref{f42}) is the analogue 
for quantum maps of (\ref{f2}).  The left-hand side 
corresponds, approximately, to $N$ times the local 
$n$-average (over a range of size of the order of 
$\varepsilon$) of $\left | \Psi_n(Q) \right |^2$, and so, 
dividing both sides by $N$ and averaging smoothly with 
respect to $Q$ (for example, taking the convolution with a 
normalized Gaussian) over a range large compared to a de Broglie 
wavelength ($\Delta Q=1$) but small compared to $N$,
\begin{equation}
\left < |\Psi_{n}(Q)|^{2} \right > \approx \frac{1}{N}+\frac{1}{N}
\rpart \sum^{\infty}_{k=1} \left< U_{Q,Q}^{k} \right >_Q\exp{(-i\theta k 
-\varepsilon k)}.
\label{f44}
\end{equation}
Here $\left< \dots \right >$ denotes a combination of the 
$n$-average and the $Q$-average $\left < \dots \right >_Q$.  

In our computations we took $\varepsilon$ large enough so that the 
dominant contributions to (\ref{f42}) and (\ref{f44}) come 
from the $k=1$ terms in the sums on the right, and so, for example, we 
may substitute (\ref{f41}) directly into (\ref{f44}).  In 
the semiclassical limit, as $N \rightarrow \infty$, the 
$Q$-average selects regions close to stationary points of 
the phase of (\ref{f41}), which we denote by $Q/N=q_f$.  
These stationary points coincide with the positions of the 
fixed points of the classical map (\ref{f40}); that is, 
$q_f$ satisfies
\begin{equation}
q_f=\frac{1}{2} \left ( j-\frac{\kappa}{2\pi}\cos (2\pi q_f) \right)
\label{f45a}
\end{equation}
for integers $j$ such that $0 \leq q_f < 1$ (see, for 
example, Boasman \& Keating 1995).  Expanding the phase of (\ref{f41})
around $q_f$ gives, up to cubic terms,
\begin{equation}
U_{Q,Q}\approx \frac{1}{\sqrt{iN}}\exp \left [ 2\pi i S_f+ \pi iN
(2-\kappa \sin (2\pi q_f))y^2 -\frac{2\pi^2 iN}{3}\kappa \cos (2\pi q_f) 
y^3 \right ],
\label{f45}
\end{equation}
where
\begin{equation}
y=\frac{Q}{N}-q_f
\label{f46}
\end{equation}
and $2\pi S_f$ denotes the phase evaluated at $q_f$.

Provided that $2-\kappa \sin (2\pi q_f) \ne 0$, this approximation is 
dominated by
the quadratic term in the exponent when $y$ is small.  It thus describes
complex-Gaussian fringes around the classical fixed points with a length-scale
(in terms of $y$)
of the order of $N^{-1/2}$.  These are the analogues of Bogomolny's fringes.
They will be resolved if the local $Q$-average is over a range that is small
compared to $N^{1/2}$ (but which still grows as $N \rightarrow \infty$).

For the example being considered here, when $\kappa < \kappa^{*} \approx 
5.94338$ 
the two values of $j$ in (\ref{f45a}), $j=0$ and $j=1$, each give rise 
to a single unstable fixed
point for which the condition $2-\kappa \sin (2\pi q_f) \ne 0$ is satisfied.
In Figure 1, we plot the left hand side of (\ref{f42}) when $\kappa=3$, with
$\varepsilon=2.2$ and for $N=1597$.  The structure around the fixed points
is clearly visible, and is most easily seen by applying the local $Q$-average
(in this case, making a convolution with a normalized Gaussian of width 
$0.02$).  

It is at bifurcations that $2-\kappa \sin (2\pi q_f) = 0$.  Then the
quadratic term in (\ref{f45}) vanishes, and the fringe structure comes instead
from the cubic term.  It thus has a $y$-length-scale of the order of 
$N^{-1/3}$.  The amplitude is the
same as in the case of isolated fixed points ($N^{-1/2}$ in the contribution to
$\left < |\Psi_{n}(Q)|^{2} \right >$).  In the language of Section 2,
this corresponds to a codimension-one bifurcation of a periodic orbit with
$r=1$ (a tangent bifurcation).

In our example, the first bifurcation occurs when $\kappa=\kappa^{*}$.  At this
parameter value, two new degenerate solutions of (\ref{f45a}) 
appear,  for both $j=0$ and $j=1$, corresponding to the birth of a 
pair of fixed points, one stable and 
the other unstable.  In Figure 2 we plot the left hand side of (\ref{f42}) with
$\varepsilon=2.2$ and $N=1597$, as above, but now with $\kappa=\kappa^{*}$.
It is apparent that the scars around the two bifurcations, at positions 
$q=0.05$ and $q=0.44$, are wider than those around the two non-bifurcating
fixed points, at positions $q=0.69$ and $q=0.81$, and that 
they are also wider than those shown in Figure 1.  
In Figure 3 we plot the left hand side of 
(\ref{f42}) close to the bifurcation point at $q=0.44$, together with the
approximation (\ref{f45}), which clearly captures the details of
the associated fringe structure.

It is straightforward now to deduce the semiclassical 
scaling with $N$ of the moments
\begin{equation}
C_{2m}(N)=N^{2m}\sum_{Q=1}^N 
\left ( \left < |\Psi_{n}(Q) |^{2} \right > - \frac{1}{N} \right )^{2m}.
\label{f47}
\end{equation}
The arguments of Section 3 suggest that 
$C_{2m}$ is of the order of $N^{-m+w}$, where $w=1/2$ away from 
bifurcations and $w=2/3$ at the bifurcation ($w$ is one plus the width 
exponents deduced from (\ref{f45}), because those were for $y$ rather than
$Ny$, as we need here).  In Figure 4 we plot $\log C_{2}$ against $\log N$ when
$\kappa=3$.  The fact that the points lie on a straight line confirms that
there is a power-law scaling; furthermore, the gradient 
is close to $-1/2$, as expected.  In
Figure 5, we make the same plot for $\kappa=\kappa^{*}$.  In this case the
gradient is close to the expected value of $-1/3$ (a possible explanation for 
the deviation is that for the range of values of $N$ shown, the 
bifurcation exponent is
contaminated by the contributions from the non-bifurcating periodic orbits).
Finally, in Figure 6 we plot
\begin{equation}
g(m)=\lim_{N\rightarrow \infty}\frac{\log{C_{2m}(N)}}{\log N},
\end{equation}
calculated numerically from the gradients of best-fitting straight lines
to plots like those in Figures 4 and 5.  For both $\kappa=3$ and $\kappa=
\kappa^{*}$ the results are in accord with the
scaling law suggested above. 

We emphasize that these numerical computations illustrate the influence of
one individual bifurcation only.  They do not test the competition which
would result from averaging over a parameter range that contains many 
different generic bifurcations, and which the analysis of Section 3 suggests
has a universal outcome for the moment exponents.

\section{Acknowledgements}

It is a pleasure to acknowledge stimulating discussions with Arnd B\"acker,
John Hannay and Jens Marklof, and comments on the manuscript by Sir Michael
Berry.  SDP wishes to thank BRIMS for financial support, and 
BRIMS and the School of Mathematics at the University of Bristol for 
hospitality during the period when this work was carried out.


\ \

\noindent {\Large References}

\ \

\noindent Agam, O. \& Fishman, S. 1994 Semiclassical criterion for 
scars in wave-functions of chaotic systems.  {\em Phys. Rev. Lett} 
{\bf 73}, 806-809.

\ \ 

\noindent Bas\'ilio de Matos, M. \& Ozorio de Almeida, A.M. 1995 Quantization
of Anosov maps.  {\em Ann. Phys.} {\bf 237}, 46-65.

\ \

\noindent Arnold, V.I. 1978 {\em Mathematical Methods in Classical
Mechanics.} Springer.
 
\ \

\noindent B\"acker, A., Schubert, R. \& Stifter, P. 1998 Rate of quantum 
ergocity in Euclidean billiards.  {\em Phys. Rev.} E {\bf 57}, 5425-5447.

\ \

\noindent Berry, M.V. 1977 Focusing and twinkling: critical exponents from
catastrophes in non-Gaussian random short waves.  {\em J. Phys.} A {\bf 10},
2061-2081.

\ \

\noindent Berry, M.V. 1983 Semiclassical mechanics of regular and irregular 
motion.  In {\em Les Houches Lecture Series} 
 (ed. G. Iooss, R.H.G. Helleman \& R. Stora), vol. 36, pp. 171-271 
Amsterdam: North Holland.

\ \

\noindent Berry, M.V. 1989, Quantum scars of classical closed orbits in phase 
space. {\em Proc. R. Soc. Lond.} A  {\bf 243}, 219-231.

\ \

\noindent Berry, M.V. 2000 Spectral twinkling. {\em New Directions in
Quantum Chaos.} Proceedings of the International School of Physics
``Enrico Femi'', 45-63. Italian Physical Society.

\ \

\noindent Berry, M.V., Hannay, J.H. \& Ozorio de Almeida, A.M. 1983 
Intensity moments of semiclassical wavefunctions.  {\em Physica} D {\bf 8},
229-242.

\ \
 
\noindent Berry, M.V,  Keating, J.P \& Prado, S.D. 1988 Orbit bifurcations and
spectral statistics. {\em J. Phys.} A {\bf 31}, L245-254.

\ \

\noindent Berry, M.V., Keating, J.P. \& Schomerus, H. 2000 Universal twinkling
exponents for spectral fluctuations associated with mixed chaology. 
{\em Proc. R. Soc. Lond.} A {\bf 456}, 1659-1668.

\ \

\noindent Berry, M.V. \& Tabor, M. 1976 Closed orbits and the regular bound 
spectrum. {\em Proc. R. Soc. Lond.} A {\bf 349}, 101-123.

\ \

\noindent Boasman, P.A. \& Keating, J.P. 1995 Semiclassical asymptotics of
perturbed cat maps.  {\em Proc. R. Soc. Lond.} A {\bf 449}, 629-653.

\ \

\noindent Bogomolny, E.B. 1988 Smoothed wavefunctions of chaotic quantum systems.
{\em Physica} D {\bf 31}, 169-189.

\ \

\noindent Colin de Verdi\`ere, Y. 1985 Ergodicit\'e et fonctions propres du 
laplacien.  {\em Commun. Math. Phys.} {\bf 102}, 497-502.

\ \

\noindent Eckhardt, B., Fishman, S., Keating, J.P., Agam, O., Main, J., 
\& M\"uller, K. 1995 Approach to ergodicity in quantum wave functions.  
{\em Phys. Rev.} E {\bf 52}, 5893-5903.

\ \

\noindent Fishman, S., Georgeot, B. \& Prange, R. E. 1996 
Fredholm method for scars.  {\em J. Phys.} A {\bf 29}, 919-937.

\ \

\noindent Gutzwiller, M.C. 1971 Periodic orbits and classical quantization 
conditions. {\em J. Math. Phys.} {\bf 12}, 343-358.

\ \

\noindent Gutzwiller, M.C. 1990 {\em Chaos in Classical and Quantum Mechanics} (New York: Springer).

\ \

\noindent Hannay, J.H. \& Berry, M.V. 1980 Quantization of linear maps on
the torus -- Fresnel diffraction by a periodic grating.  {\em Physica} D
{\bf 1}, 267-290.

\ \

\noindent Heller, E.J. 1984 Bound state eigenfunctions of classically 
chaotic Hamiltonian systems - scars of periodic orbits.  
{\em Phys. Rev. Lett.} {\bf 53}, 1515-1518.

\ \

\noindent Kaplan, L. 1999 Scars in quantum chaotic wavefunctions.  
{\em Nonlinearity} {\bf 12}, R1-R40.

\ \

\noindent Keating, J.P. 1991 The cat maps: quantum mechanics and classical
motion.  {\em Nonlinearity} {\bf 4}, 309-341.

\ \

\noindent Keating, J.P., Mezzadri, F. \& Robbins, J.M. 1999 Quantum 
boundary conditions for torus maps.  {\em Nonlinearity} {\bf 12}, 579-591.

\ \

\noindent McDonald, S.W. 1983 {\em Lawrence Berkeley Laboratory Report} LBL -
 14837.

\ \

\noindent Meyer, K.R. 1970 Generic bifurcations of periodic points.
{\em Trans. Am. Math. Soc.} {\bf 149}, 95-107.

\ \

\noindent Meyer, K.R. 1986 Bibliographic notes on generic bifurcations
in Hamiltonian Systems.
{\em Contemp. Math.} {\bf 56}, 373-381.

\ \

\noindent Ozorio de Almeida, A.M. 1988 {\em Hamiltonian Systems: Chaos
and Quantization.} Cambridge University Press.

\ \

\noindent Ozorio de Almeida, A.M. \& Hannay, J.H. 1987 Resonant periodic orbits
and the semiclassical energy spectrum. {\em J. Phys} A {\bf 20}, 
5873-5883.

\ \

\noindent Percival, I.C. 1973 Regular and irregular spectra.  
{\em J. Phys.} B {\bf 6}, L229-L232.

\ \

\noindent Schomerus, H. 1998 Periodic orbits near bifurcations of codimension
two: Classical mechanics, semiclassics and Stokes transitions. {\em
J. Phys.} A {\bf 31}, 4167-4196.

\ \

\noindent Schomerus, H. \& Sieber, M. 1997 Bifurcations of periodic orbits
and uniform approximations. {\em J. Phys.} A {\bf 30}, 4537-4562.

\ \

\noindent Shnirelman, A.I. 1974 Ergodic properties of eigenfunctions (in 
Russian).  {\em Usp. Math. Nauk} {\bf 29}, 181-182.

\ \

\noindent Sieber, M. 1996 Uniform approximation for bifurcations of 
periodic orbits
with high repetition numbers. {\em J. Phys.} A {\bf 29}, 4715-4732. 

\ \

\noindent Sieber, M. \& Schomerus, H. 1998 Uniform approximations for 
period-quadrupling bifurcations. {\em J. Phys.} A {\bf 31}, 165-183.

\ \

\noindent Tomsovic, S., Grinberg, M. \& Ullmo, D. 1995 Semiclassical trace 
formulas of near-integrable systems: Resonances. {\em Phys. Rev. Lett.}
{\bf 75}, 4346-4349.

\ \

\noindent Ullmo, D., Grinberg, M. \& Tomsovic, S. 1996 Near-integrable systems:
Resonances and semiclassical trace formulas. {\em Phys. Rev.} E 
{\bf 54}, 136-152.

\ \

\noindent Zelditch, S. 1987 Uniform distribution of eigenfunctions on compact 
hyperbolic surfaces.  {\em Duke Math. J.} {\bf 55}, 919-941.


\newpage 
 
\centerline{FIGURES}

\begin{figure}[htb]
  \begin{center}
   \leavevmode
   \psfig{figure=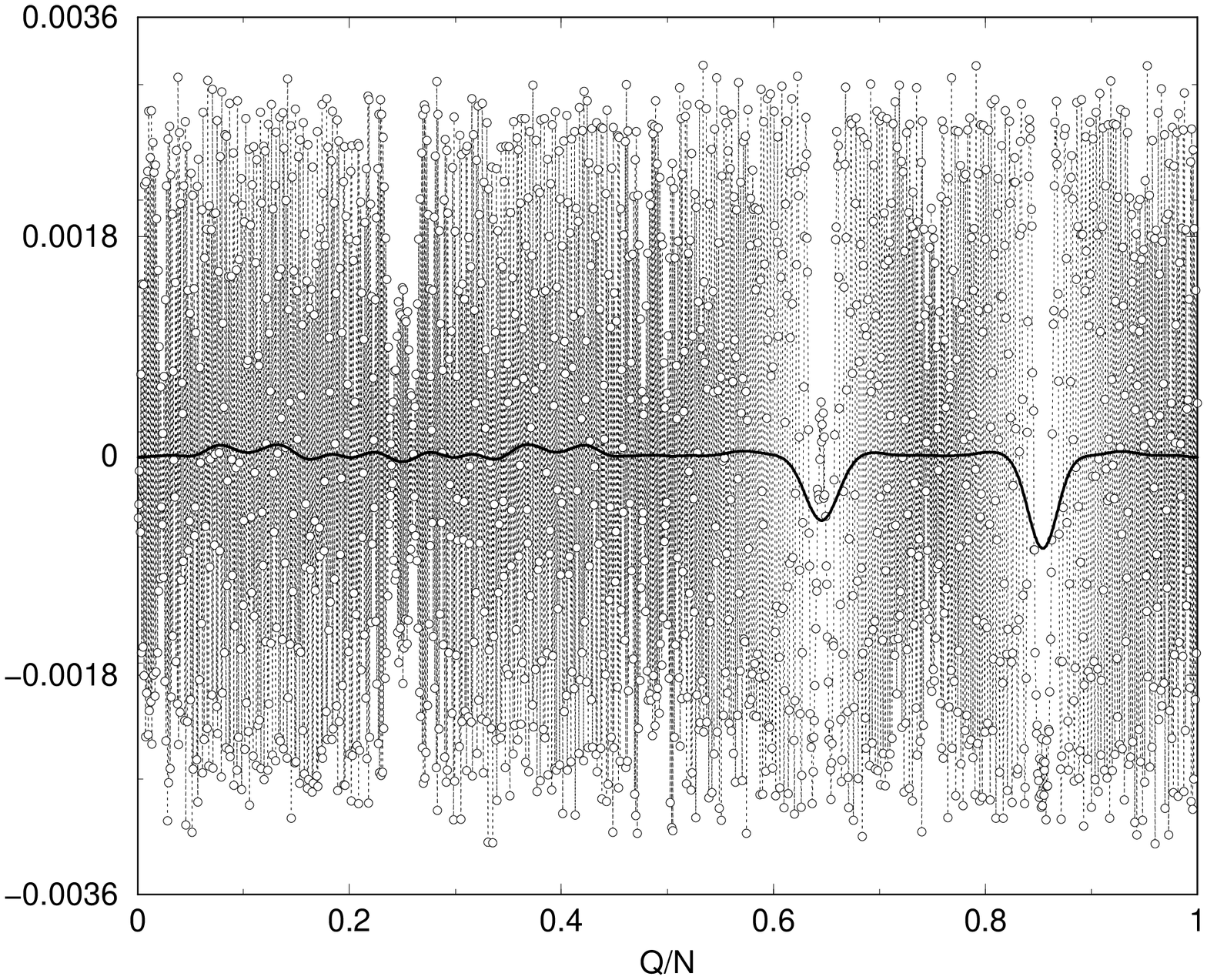,width=16.0cm,angle=270}
  \end{center}
\caption{$\sum_{n}\left | \psi_{n}(Q) \right |^{2}\delta_{\varepsilon}
\left (\theta-\theta_{n} \right )-1$, with $\varepsilon=2.2$, $\kappa=3$
and $N=1597$ (circles connected by dotted lines).  Also shown is a
convolution of the data with a normalized Gaussian of width 0.02 (bold line).
The positions of the fixed points are $q=0.65$ and $q=0.85$.}
\label{fig:k3}
\end{figure}

\begin{figure}[htb]
  \begin{center}
   \leavevmode
   \psfig{figure=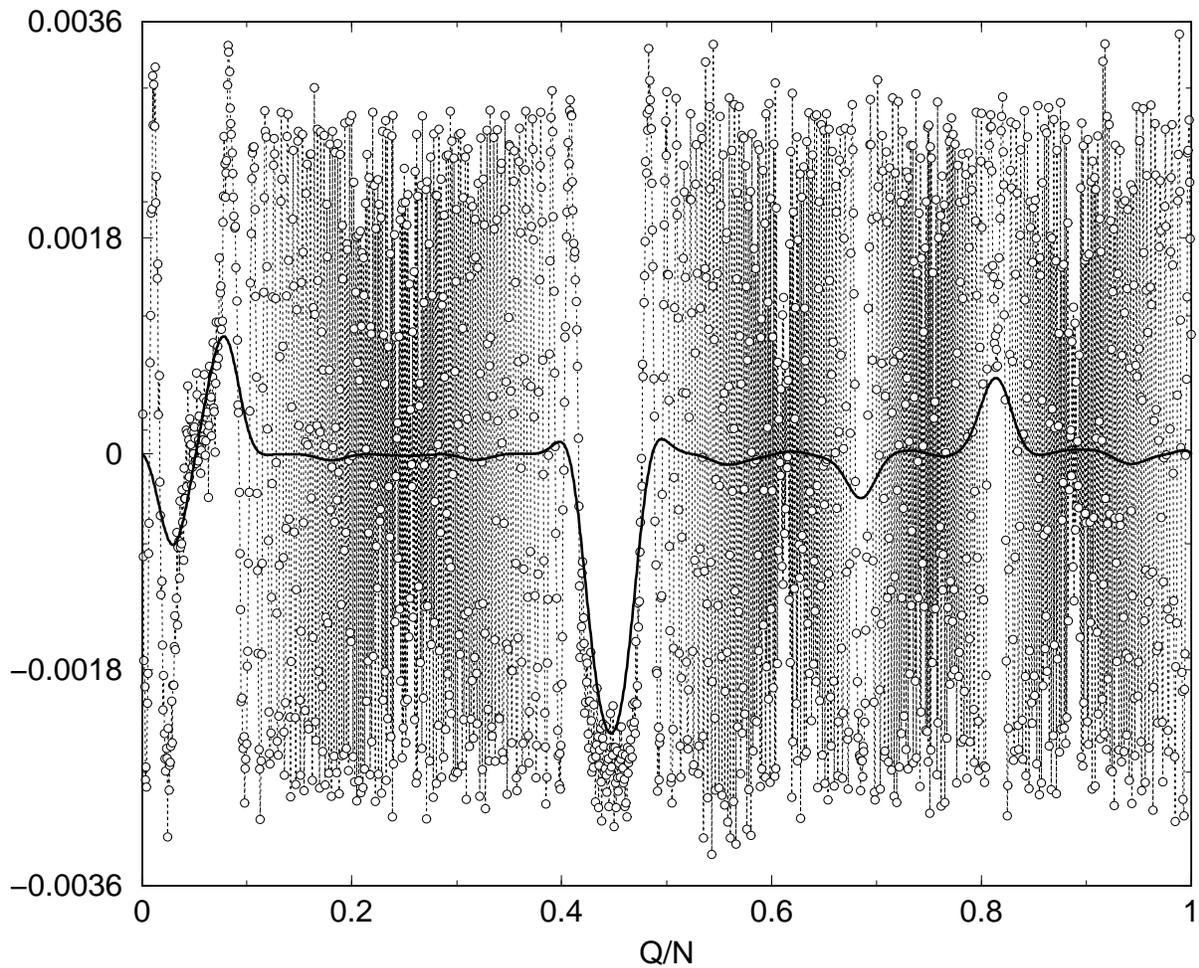,width=16.0cm,angle=270}
  \end{center}
\caption{$\sum_{n}\left | \psi_{n}({Q}) \right |^{2}\delta_{\varepsilon}
\left (\theta-\theta_{n} \right )-1$, with 
$\varepsilon=2.2$, $\kappa=\kappa^{*}$
and $N=1597$ (circles connected by dotted lines).  Also shown is a
convolution of the data with a normalized Gaussian of width 0.02 (bold line).
There are unstable fixed points at $q=0.69$ and $q=0.81$, and bifurcations
at $q=0.05$ and $q=0.44$.}
\label{fig:kbif}
\end{figure}

\begin{figure}[htb]
  \begin{center}
   \leavevmode
   \psfig{figure=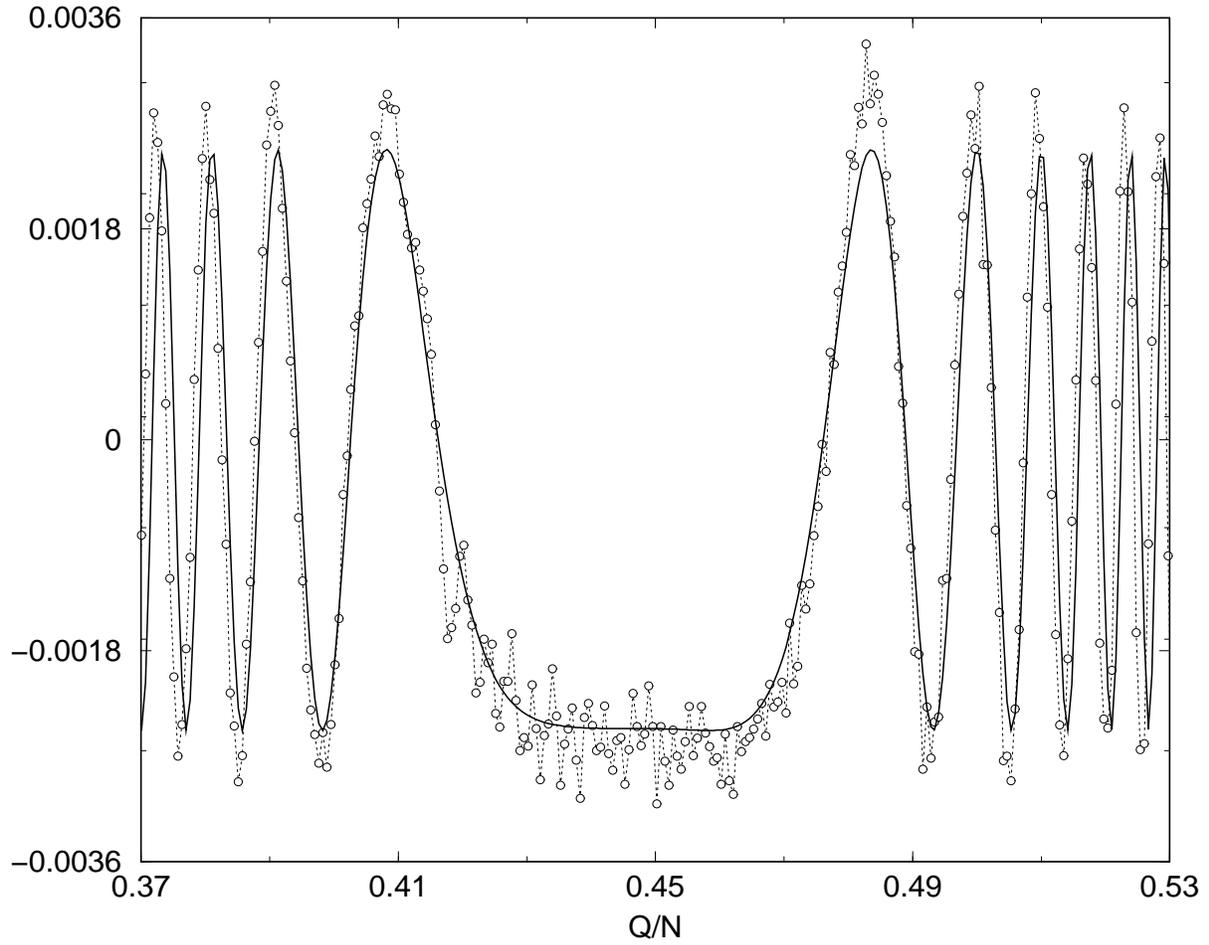,width=16.0cm,angle=270}
  \end{center}
\caption{$\sum_{n}\left | \psi_{n}({Q}) \right |^{2}\delta_{\varepsilon}
\left (\theta-\theta_{n} \right )-1$, with $\varepsilon=2.2$, $\kappa=\kappa^{*}$
and $N=1597$, as in Figure 2, in the neighbourhood of the bifurcation at 
$q=0.44$ (circles connected by dotted lines).  Also shown is the local 
approximation obtained by substituting (\ref{f45}) into (\ref{f42}) 
(bold line).}
\label{fig:local}
\end{figure}

\begin{figure}[htb]
  \begin{center}
   \leavevmode
   \psfig{figure=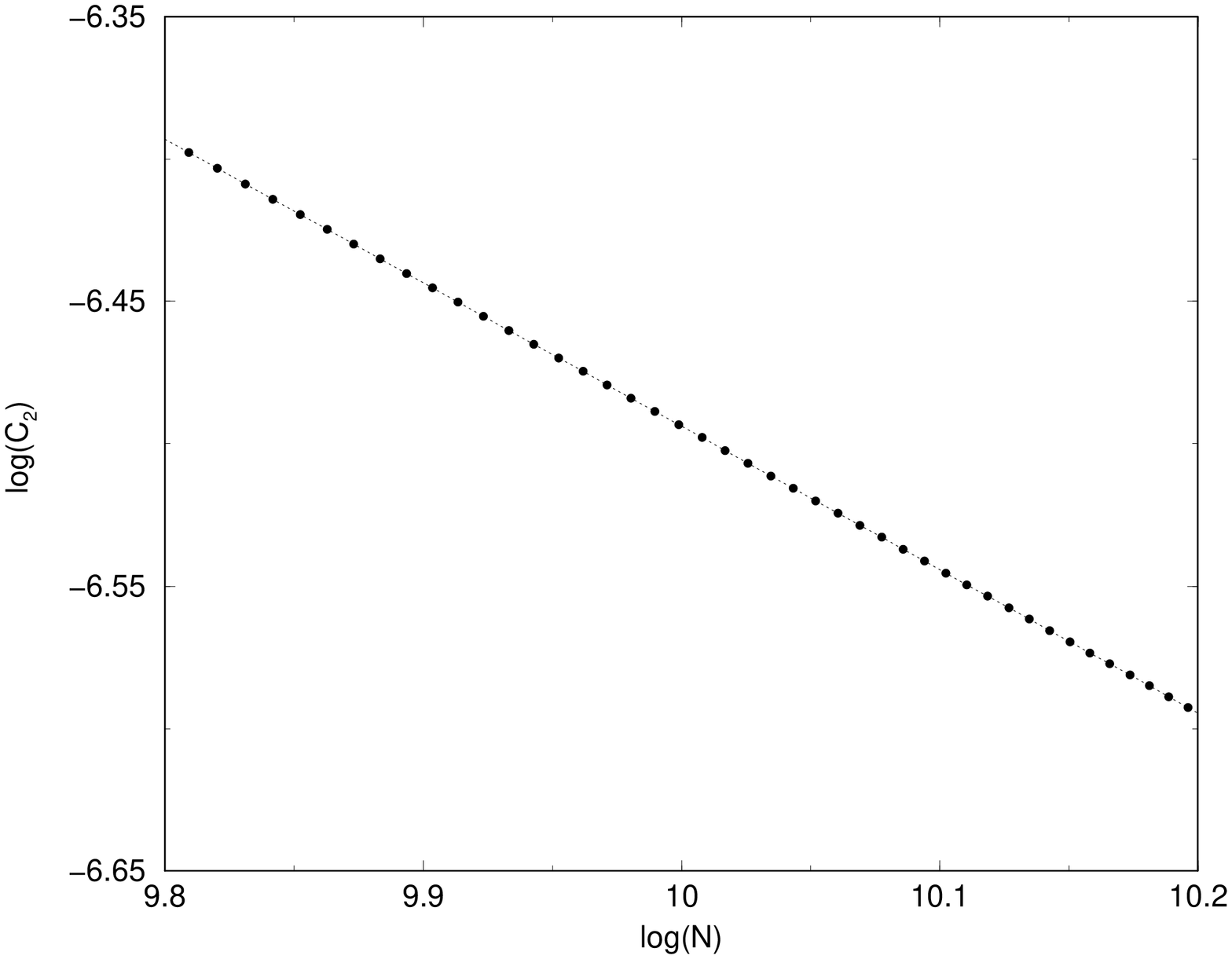,width=16.0cm,angle=270}
  \end{center}
\caption{$\log C_2$, calculated using a local $Q$-average of size 
$0.02N^{1/2}$, 
plotted against $\log N$ when $\kappa=3$ (circles). Also shown is a 
best-fitting straight line, which has gradient -0.50.}
\label{fig:slope3}
\end{figure}

\begin{figure}[htb]
  \begin{center}
   \leavevmode
   \psfig{figure=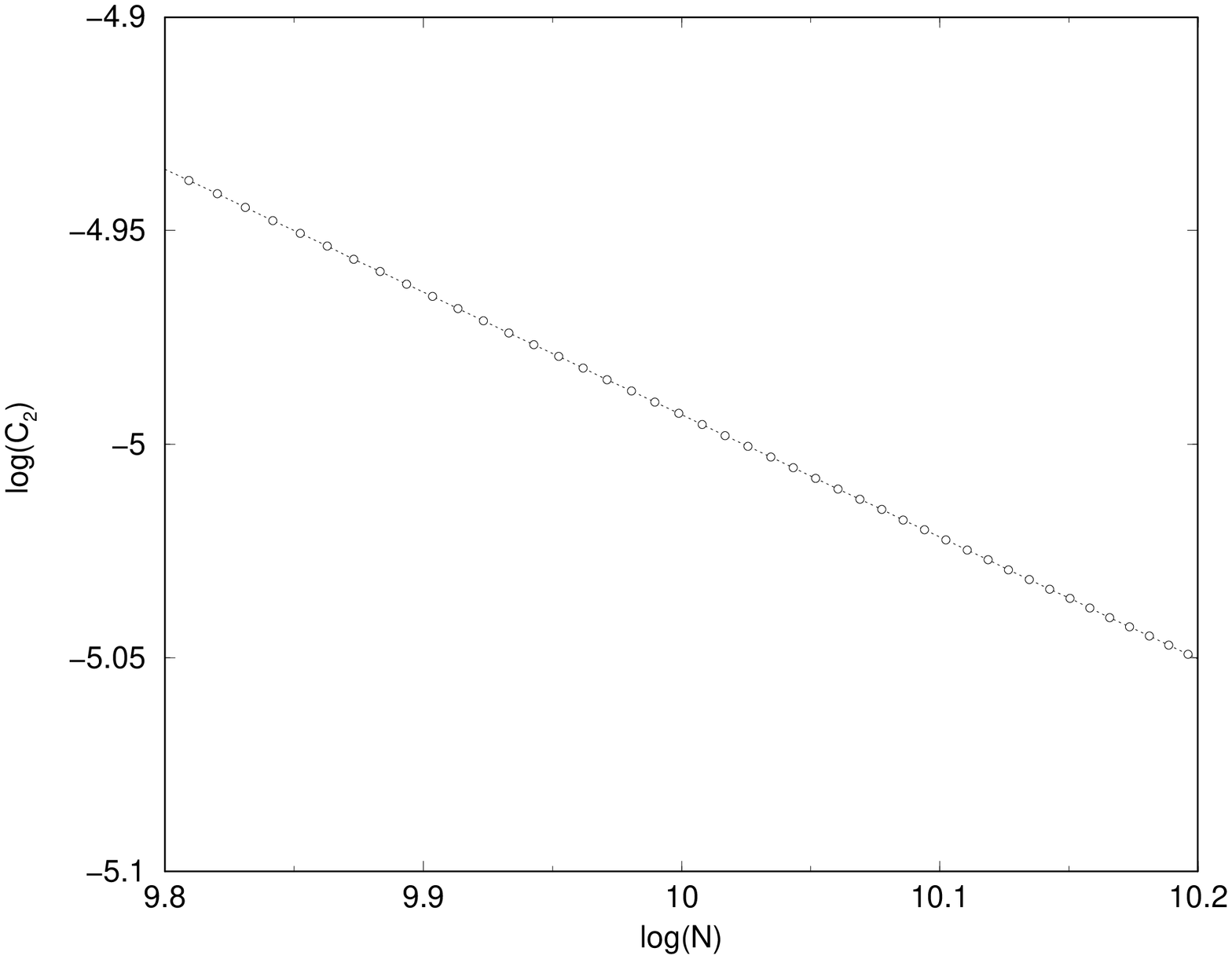,width=16.0cm,angle=270}
  \end{center}
\caption{$\log C_2$, calculated using a local $Q$-average of size 
$0.02N^{1/2}$, 
plotted against $\log N$ when $\kappa=\kappa^{*}$ (circles). Also shown is a 
best-fitting straight line, which has gradient -0.29.}
\label{fig:slopebif}
\end{figure}

\begin{figure}[htb]
  \begin{center}
   \leavevmode
   \psfig{figure=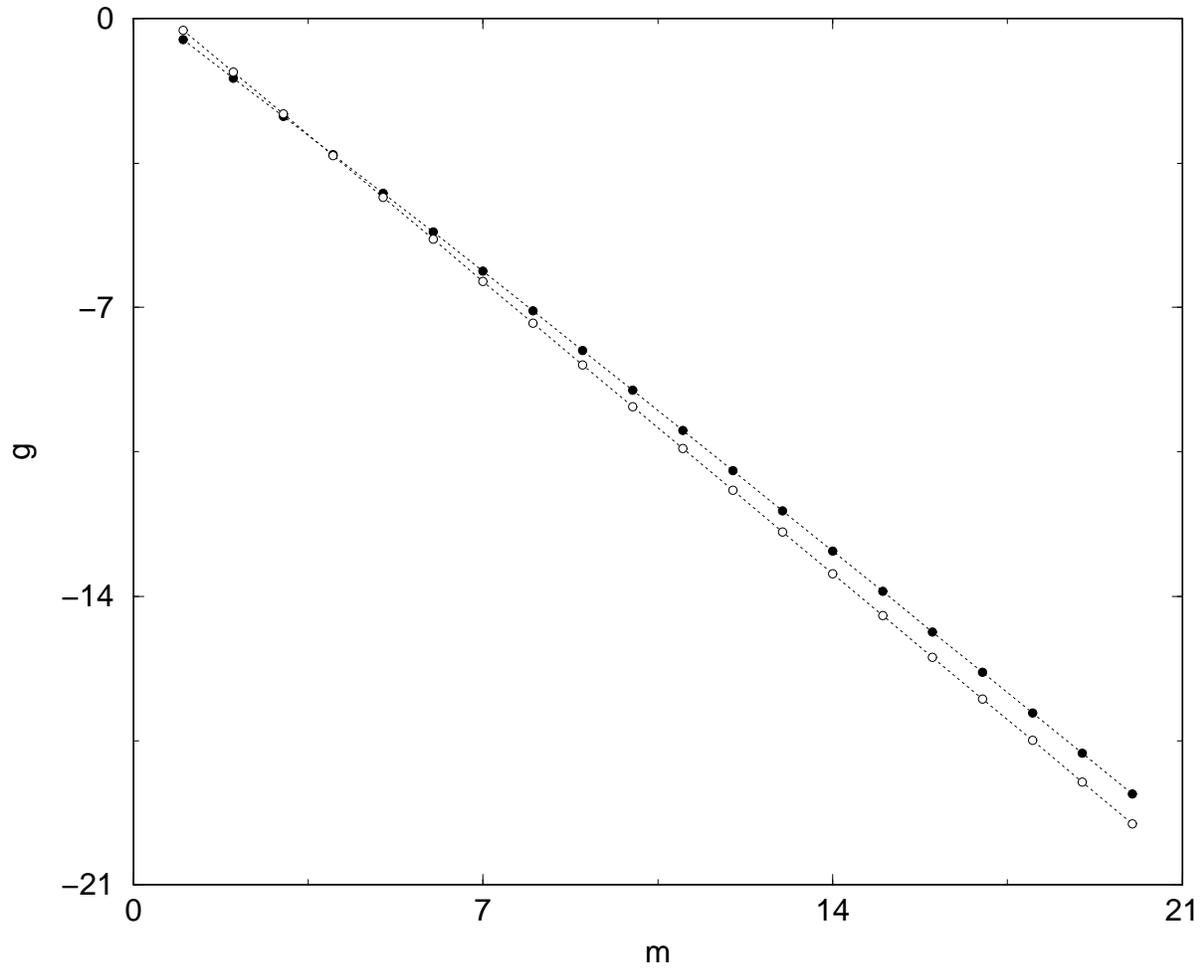,width=16.0cm,angle=270}
  \end{center}
\caption{$g(m)$ plotted against $m$, for $\kappa=3$ (full circles), and
$\kappa=\kappa^{*}$ (open circles).  Also shown are the best fitting straight
lines: $g=-0.96m+0.57$ and $g=-1.01m+0.72$, respectively.}
\label{fig:higherm}
\end{figure}

\end{document}